\documentstyle[epsf]{article}

\newcommand{\be}{\begin{equation}}
\newcommand{\ee}{\end{equation}}

\begin{document}

\vspace{5mm}
\begin{center}
{\LARGE \bf The $\eta$ and $\eta'$ mesons from lattice QCD.}\\[10mm] 
 
 {\bf   C.~Michael\footnote{email: c.michael@liv.ac.uk} \\ 
Theoretical Physics Division, Dept. of Mathematical Sciences, 
          University of Liverpool, Liverpool L69 3BX, UK }\\[2mm]


\end{center}

\begin{abstract}

 Lattice QCD allows a first-principles study of QCD with the freedom  to
vary the number and masses of the quarks. I present results on the 
flavour singlet correlations (this illuminates OZI violating effects)
for  mesons. Concentrating on the pseudoscalar mesons, the flavour
singlet  mass splitting ($\eta$, $\eta'$ mass splitting) appears
naturally.
  I also present results on an investigation of decay constants for the
$\eta$ and $\eta'$ ($f_{\eta}$)  and discuss which quantities may be
accessible  in future lattice studies.
 The Witten-Veneziano  approach can also be explored by  determining the
quenched topological susceptibility on a lattice.

\end{abstract}
%


\section{Introduction}

 There is considerable interest in understanding hadronic decays
involving  $\eta$ and $\eta'$ in the final state. The phenomenological
study of hadronic processes involving flavour  singlet pseudoscalar
mesons makes assumptions about their composition. Here I  address the
issue of the nature of the $\eta$ and $\eta'$ from QCD directly,  making
use of lattice techniques. 

 Lattice QCD directly provides a bridge between the underlying quark
description and  the non-perturbative hadrons observed in experiment. 
The amplitudes to create a given meson from the vacuum with a particular
 operator made from quark fields are measurable, an example being  the
determination of $f_{\pi}$. It also allows a quantitative study of the
disconnected quark contributions that arise in the flavour singlet
sector.   The lattice approach  provides other information such as that
obtained by varying the number of quark flavours and their masses. 

In the case of pseudoscalar mesons, the chiral perturbation theory
approach  also provides links between a quark description and the
hadronic states. Indeed since lattice studies are increasingly difficult
as  the quark mass is reduced, the objective is to find a region of
quark mass  which is accessible from the lattice and for which chiral
perturbation  theory is reasonably convergent. As an example, lattice
studies~\cite{chptukqcd} can be used to estimate  the size of the higher
order terms in a chiral lagrangian approach.

 For the pion, lowest order chiral perturbation theory has the  well
known consequence  that the  decay constant $f_{\pi}$ describes
quantitatively both the $\mu \nu$  and $\gamma \gamma$ decays. For the
flavour singlet states ($\eta$ and $\eta'$), the situation is more
complicated~\cite{chpt}.  The axial anomaly  now involves a gluonic
component and the definition of decay constants is  not straightforward.
From the chiral perturbation theory description, one expects the mixing
of $\eta$ and $\eta'$ to be most simply  described in a quark model
basis.
 In the flavour singlet sector, for pseudoscalar mesons, one then has
contributions  to the mass squared matrix with quark model content $(u
\bar{u} +d \bar{d})/\sqrt{2}$  and $s \bar{s}$ (which are labelled as $n$
and $s$ respectively):

 \be   
   \left( \begin{array}{cc} m_{nn}^2 +2x_{nn} &  \sqrt{2} x_{ns} \\
           \sqrt{2} x_{ns} &  m_{ss}^2+x_{ss}  \end{array} \right)
  \ee

 Here $m$ corresponds to the mass of the flavour non-singlet  eigenstate
and is  the contribution to the mass coming from connected fermion
diagrams while $x$ corresponds to the contribution from disconnected
fermion diagrams. In the limit of no mixing (all $x=0$, the OZI
suppressed case), then   the quenched QCD  result is that the $\eta$
is degenerate with the $\pi$ meson and the $\eta'$ would correspond to
the $s \bar{s}$ pseudoscalar meson. This is not  the case, of course,
and the mixing contributions $x$ are important.

\begin{figure}[h]

\epsfxsize=10cm\epsfbox{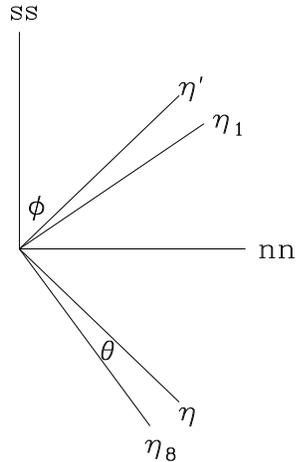}

 \caption{ The mixing illustrated in the quark model basis
for flavour singlet pseudoscalar mesons. 
 }
 \label{mixing}

\end{figure}

\begin{figure}[h]

\epsfxsize=10cm\epsfbox{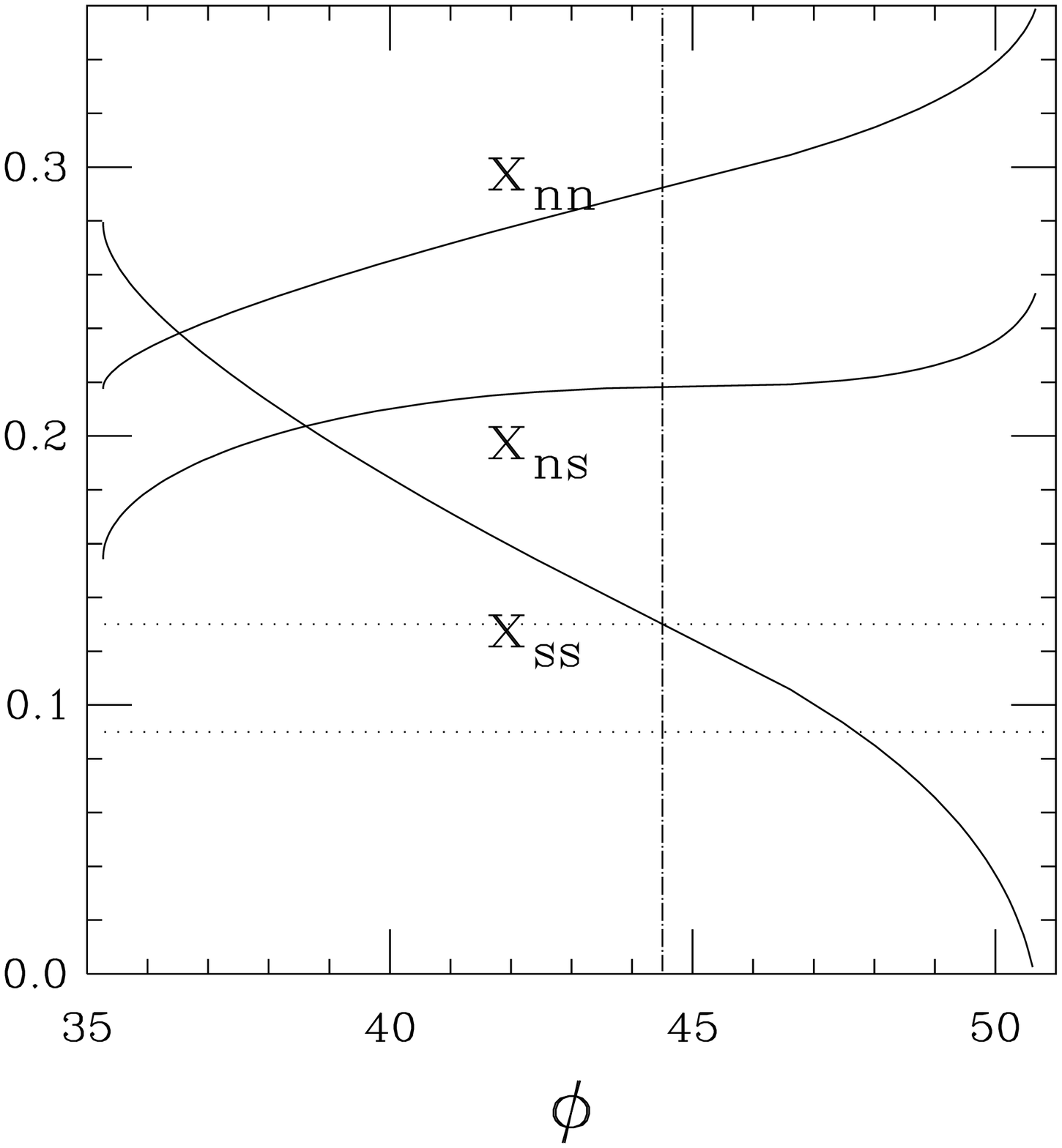}

 \caption{ The mass mixing parameters $x$ in GeV$^2$ versus $\eta$,
$\eta'$ mixing angle $\phi$ in the $\eta_{nn},\ \eta_{ss}$ basis.  The
horizontal dotted lines give the allowed range from the lattice
determination  of $x_{ss}$. The vertical line  illustrates the preferred
solution.
 }
 \label{xxx}

\end{figure}

Using as input $m_{nn}$, $m_{ss}$, $m_{\eta}$ and $m_{\eta'}$, the three
mixing  parameters $x$ cannot be fully determined. It is usual to
express the resulting one parameter freedom  in terms of a mixing angle,
as illustrated in fig.~1, here defined by
 \be
 \eta =\eta_{nn} \cos \phi - \eta_{ss} \sin \phi \ \ \ 
 \eta'=\eta_{nn} \sin \phi + \eta_{ss} \cos \phi
 \ee 
  The resulting values of the mixing parameters $x$ are shown in  fig.~2
(the  input value for $m_{ss}$ will be discussed later).

 The $\eta$ and  $\eta'$ mesons are often described in an SU(3) 
motivated quark basis, namely
$\eta_8=(u\bar{u}+d\bar{d}-2s\bar{s})/\sqrt{6}$, 
$\eta_1=(u\bar{u}+d\bar{d}+s\bar{s})/\sqrt{3}$.  The mixing angle
$\theta$  in this basis (see fig.~1) would be given by $\phi-54.7^0$ in
a lowest order chiral perturbation theory. In order to have $f_K \ne
f_{\pi}$, one needs  higher order terms in the chiral perturbation
theory treatment and then the mixing scheme  becomes more
complicated~\cite{chpt} in this basis with more than one angle needed.

 In the SU(3) symmetric limit,  $m_{nn}=m_{ss}=m$ and
$x_{nn}=x_{ns}=x_{ss}=x$, so that  only one mixing parameter is
relevant and the mixing matrix simplifies  considerably to a diagonal
form with elements $m^2$ (octet) and $m^2+3x$ (singlet).  Previous
lattice studies~\cite{kuramashi} have  used degenerate quarks, so have
explored this case and have found that the mixing parameter $x$ is of a
magnitude which  can explain qualitatively the observed splitting
between the $\eta$ and $\eta'$ mesons.

 Here I describe a  non-perturbative  study in QCD from first
principles which will be able to establish the values of the mixing
parameters $x$, including the pattern of SU(3) breaking.
 This more comprehensive study would take into account the different
masses of the light ($u$ and $d$) quarks and the heavier $s$ quark.
Within the lattice approach, it is not at present feasible to evaluate
using quarks as  light as the nearly massless $u$ and $d$ quarks and
also it is more tractable to use  an even number of degenerate quarks in
the vacuum.  As I shall show, despite these restrictions,  a thorough
study of the mixing between  $\eta$ and $\eta'$ is possible.

 I focus here on the results of lattice evaluations and  I address four
topics where lattice input permits us to construct a  firm foundation
for the $\eta$, $\eta'$ mixing:

 \begin{itemize}
 \item From comparing pseudoscalar meson masses with valence quarks of two
different masses (namely meson masses $m_{11}$, $m_{12}$ and $m_{22}$),
one can estimate the  mass $m_{ss}$ of the unmixed $\bar{s} s$ meson,
given the  observed $m_{ns}$ and $m_{nn}$ masses (ie $K$ and $\pi$
respectively).

 \item From measuring the mixing parameters $x_{11}$, $x_{12}$ and
$x_{22}$  between initial and final flavour singlet states  consisting
of either quark 1 or 2 with different masses, one can  establish
consistently the pattern of SU(3) breaking in the mixing: obtaining 
a mixing angle close to $45^0$ in the quark model basis.

 \item For $N_f=2$ degenerate flavours of quark, one can  determine  the
pseudoscalar decay constants for the flavour singlet  and non-singlet 
meson. This input allows us to discuss the relation between  the
observed $\gamma \gamma$ decay modes of $\pi^0$, $\eta$  and $\eta'$ and
the underlying quark content. 

\item The topological susceptibility can be measured in both quenched and 
full QCD and shows the behaviour expected for it to be related to 
the flavour singlet pseudoscalar mass generation.

 \end{itemize}

\section{Connected and disconnected contributions}

\begin{figure}[h]

\begin{center}
\epsfxsize=6cm\epsfbox{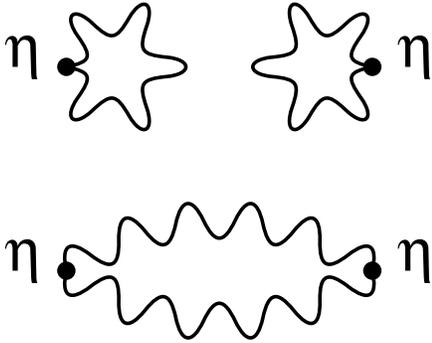}

 \caption{ The  disconnected (above) and connected (below) correlators 
for flavour singlet pseudoscalar mesons at time separation $t$. Here 
the wiggly lines represent full quark propagators.
 }
 \label{dbyc}

 \end{center}
\end{figure}

 On a lattice one can create  pseudoscalar mesons by using any operator
that has the required quantum numbers.  The  simplest choice is
$\bar{\psi} \gamma_5 \psi$ where $\psi$ represents  a quark field. A
suitable  sum over quark flavours yields the required state ($\pi$,
$\eta$, etc).  By summing over spatial positions, one can create a state
with zero momentum.  One then  studies these pseudoscalar mesons by
creating them at  time $t=0$ and annihilating them at time $t$  and
using the full quark propagators evaluated on the lattice to construct
the correlator ${\cal C}(t)$. The basic strategy is then that  at large
$t$ where excited state contributions will be negligible, this
correlator behaves as $e^{-mt}$ where $m$ is the lightest mass
 state with the specified quantum numbers. Hence one can determine meson
 masses.

 A feature that occurs exclusively for flavour singlet mesons, is that 
there are two independent ways to link up the quarks and antiquarks 
between the initial and final state - as illustrated in fig.~3.
Moreover, these connected and disconnected contributions  can be
evaluated separately on a lattice, so giving  additional insight. 
Indeed the connected contribution $C(t)$ is just the same for flavour
singlet and non-singlet mesons. Thus the disconnected contribution $D(t)$ 
is responsible for the difference between them. So, at large $t$ one has 
for the flavour non-singlet correlator
 \be
 {\cal C}(t)= C(t) \to c e^{-mt}
 \ee
 and for the flavour singlet correlator
 \be
 {\cal C}(t)= C(t)+D(t) \to d e^{-m't}
 \ee
 where $m$ is the flavour non-singlet mass and $m'$ is the flavour
singlet  mass.
 
 Note that this implies that at large $t$
 \be
  {D(t) \over C(t)}=  {d \over c} e^{-(m'-m)t} -1 \approx 
      {d \over c}-1 -  {d \over c}(m'-m)t +O[(m'-m)^2 t^2]
 \ee

 Thus a study of $D/C$ gives information on the mass splitting between 
singlet and non-singlet.  This has been explored on a  lattice for
mesons of all $J^{PC}$ values that can be made by local quark
bilinears~\cite{discall}. This study finds that the only sizeable
contributions  to $D$ come for the scalar and pseudoscalar quantum
numbers. This conclusion  is in agreement with the experimental
observation that the meson spectrum  is approximately ideal (ie as in
the naive quark model) except for those  two quantum numbers. 

 For the pseudoscalar case the behaviour of $D/C$ is such that the 
flavour singlet meson is heavier than the flavour non-singlet, as indeed 
required by experiment. Also, for $N_f$ degenerate flavours of quark, 
the flavour singlet mass is given by $m'=\sqrt{m^2+N_f x}$, in terms of the 
mass matrix approach described above. This gives us access 
to values of $x$ from the lattice for different values of $N_f$. 

 As well as the simple case where all quarks have the same mass and are 
present as sea-quarks in the vacuum, one can consider on a lattice 
other cases. One relevant case is with 2 flavours of sea quark (thought
of as $n$ quarks), but  with mesons made out of $n \bar{s}$ or $s
\bar{s}$  where the $s$ quark is heavier and so not included in the
sea. 
 Here we have a partially quenched scenario since the $s$ quarks are not 
present in the sea. In this case one can show that the disconnected diagram 
with $nn$ as a source and $ss$ as a sink can be expressed~\cite{cmcmeta} as 
 \be
   { D_{ns}(t) \over \sqrt{C_{nn}(t) C_{ss}(t)} } =
  { \sqrt{2} x_{ns} t\over 2 \sqrt{m_{nn} m_{ss} } } + O(x^2 t^2)
 \ee
 which allows $x_{ns}$ to be evaluated. Likewise $x_{ss}$ can be evaluated.

\section{Lattice results}

 \subsection{ The $s \bar{s}$ pseudoscalar mass}

 A meson made from  $s \bar{s}$ quarks is necessarily flavour singlet
and  so can mix with other flavour singlet operators via gluonic
interactions.  In order to understand the mixing between the $\eta$ and
$\eta'$, it is  important to ask a hypothetical question: what would be
the mass of a  pseudoscalar meson made of $s \bar{s}$ quarks but which
is flavour non-singlet? This question can be addressed readily by lattice
studies where  the extra interactions (disconnected diagrams)
associated with the flavour  singlet case can be removed explicitly.

 Chiral symmetry considerations lead to the expectation that  the
pseudoscalar meson composed of quarks of mass $M_q$ has mass squared
$m^2$  which behaves  linearly with $M_q$ at small quark mass. However,
at large quark mass ($c$ and $b$ quarks for instance), one expects the 
meson mass to vary approximately linearly with the quark mass. Here we
are not concerned with the region of  very small quark mass where chiral
logs are important~\cite{chpt}, so one can  summarise this  behaviour by

 \be
   m^2 = b M_q + c M_q^2 + O(M_q^3)
\label{cseq}
 \ee

\noindent  For a pseudoscalar meson  made of two different quarks of
mass  $M_n$ and $M_s$,  we shall assume its  mass only depends on
$(M_n+M_s)/2$ and not on $(M_s-M_n)/2$ as found in lattice
studies~\cite{ukqcd0} and in lowest order chiral perturbation theory. 
 If  eq.~\ref{cseq} were valid with just the linear term  in the quark
mass(ie $c=0$),  then  one directly obtains the required  mass of the
pseudoscalar meson composed of $s$ quarks, $m_{ss}^2=
2m_{sn}^2-m_{nn}^2$, that is $2K^2-\pi^2$, leading to $m_{ss}=0.687 $
GeV.

 This can be explored on a lattice by measuring the pseudoscalar meson
mass  for valence quarks in combinations 11, 22 and 12.  Then, for 
small $c/b$, one has  
 \be
 {c \over 4 b^2} = { {1 \over 2}(m_{11}^2 +m_{22}^2)  - m_{12}^2  \over
       (m_{22}^2 -m_{11}^2 )^2 }
 \ee
 One can then to use dynamical configurations with sea quarks of type 1
and   consider the propagation of mesons made of either quark 1 or
quark 2, where quark 2  corresponds to a heavier  quark. 

From such lattice studies,  a statistically significant curvature from
the $c$ term is obtained.  Applying  this value of $c$ to the
determination of the $m_{ss}$ mass  from the $\pi$ and K masses, 
gives~\cite{cmcmeta} a relative shift upwards due to the curvature term
($c$) of 1.1(3)\%,  corresponding to a value of $m_{ss}=0.687+0.008$
GeV.

\subsection{Flavour-singlet mixing}

 The mass splitting between flavour non-singlet and singlet mesons can
be measured, as discussed above,  using  lattice evaluation of
disconnected quark propagators. This is not an easy task:  the
contamination from excited states is difficult to remove and the 
statistical errors turn out to be relatively large. Initial studies have
been in the quenched approximation~\cite{kuramashi,gusken,MQA}. Here,
although there  is no flavour splitting of the masses,  the mass
splitting matrix element $x$  can be evaluated.  It is, however,
preferable  to be able to study the mass splitting directly and hence
here I  focus on results from full QCD simulations
~\cite{lat99,sesam,cmcmeta}. Some technical details  of lattice
simulations of disconnected correlators can be found in
ref.~\cite{gusken}.

Already in the first quenched studies~\cite{kuramashi}, it was found
that $x$  increases as the quark mass is decreased.  Moreover, it will
be interesting to check to see if there is a factorisation  of $x$ as
expected in some chiral perturbation theory descriptions~\cite{chpt},
namely $x_{12}^2=x_{11} x_{22}$.
 Consider now  the more realistic (partially quenched) 
case: with heavier quarks  of type 2 in a sea of two flavours of quarks of 
type 1.

Setting the sea quark mass to strange in both quenched and  $N_f=2$
evaluations leads to a consistent lattice estimate~\cite{cmcmeta} of
$x_{ss}$ in the range  0.09 to 0.13 GeV$^2$. 
 This value is also consistent with that reported from a study of
$N_f=2$ by the CP-PACS collaboration~\cite{cppacs} with strange quarks 
which gives  values of   $x_{ss}=0.10$ GeV$^2$ and 0.14
GeV$^2$  (depending on using $t_{\rm min}=2,\ 3$ in fits, respectively).
These lattice values are obtained at quite coarse lattice spacings
 and there may be some additional systematic error  arising from the
extrapolation to the continuum limit. The authors~\cite{cmcmeta} have,
however, chosen to use a  non-perturbatively improved  fermion
action~\cite{ukqcd}  to minimise this extrapolation error.

 They are unable to determine directly the mixing strengths $x$ for
lighter quarks than strange.
  However, the  $x$ values do show a
decrease with increasing quark mass and also  approximate factorisation.
  So they  assume that the value of $x$
continues to increase as the quark mass is decreased below strange in a
similar  way to the decrease  seen from twice strange (type 2) to
strange (type 1) where approximate factorisation is found. 

Consider now the consequence of this lattice determination of the
mixing, using input masses $m_{nn}=0.137$ GeV, $m_{ss}=0.695$ GeV (as
discussed above) and with  $x$ values in line with the results above,
namely $x_{ss} \approx 0.12$ GeV$^2$,   $x_{ns}^2 \approx x_{nn} x_{ss}$
and, though with big errors from the extrapolation, 
$x_{nn}/x_{ss} \approx 2$.
 Fig.~2  shows the $x$ values needed to  reproduce the known $\eta$
and $\eta'$ masses for each mixing angle $\phi$.
 The lattice determination of $x_{ss}$ is shown by the dotted horizontal
band.  Keeping close to this band while satisfying the other lattice
constraints  is possible for the mixing illustrated by the vertical
line. This has  $x_{nn}=0.292,\ x_{ns}=0.218, \ x_{ss}=0.13$ GeV$^2$
which gives a description of the  observed $\eta$ and $\eta'$ masses
while being consistent with the  QCD inspired evidence about the mixing
strengths. This assignment  corresponds to a mixing angle $\phi$ in the
$\eta_{nn},\ \eta_{ss}$ basis of 44.5$^0$. Note that this  is almost maximal
which  implies that the quark content (apart from the relative sign)  of
the $\eta$ and $\eta'$ meson is the same. The corresponding mixing angle
in the $\eta_8$, $\eta_1$  basis (modulo comments above) is a value of 
$\theta$ of $-10.2^0$.

\subsection{Flavour-singlet decay constants}

 One contribution to the decays of $\pi^0,\ \eta$ and $\eta'$ to $
\gamma \gamma$ is  via the quark triangle diagram. The quark model gives
a decay proportional to $Q_i^2$ for the contribution from a quark of
charge $Q_i$. Thus for the $\pi^0$ meson and the  flavour-singlet $nn$
and $ss$ mesons,  the quark charge contributions to the decay amplitudes
would be in the ratio  $1:5/3:\sqrt{2}/3$. The experimental~\cite{pdg} 
reduced decay amplitudes for $\pi^0$, $\eta$, and $\eta'$ are in the
ratio $1.0 : 1.00(10): 1.27(7)$.  This information can be used to
analyse the  quark content of the pseudoscalar mesons subject to a
quantitative understanding of the decay mechanisms.
 The traditional approach assumes that the decay constants for the
decays  of the three mesons are the same and then the relative decay
amplitudes give information on the  quark content. This suggests a
mixing angle of $\theta \approx -20^0$ is preferred~\cite{chpt,pdg}.

 For the flavour non-singlet mesons we have 
 \be
 \langle 0 | A^{\mu}| \pi \rangle = f_{\pi} q^{\mu}
 \label{pcacpi}
 \ee
 where the axial current is the local quark bilinear. From evaluating
this  expression on a lattice, the pion decay constant can be
determined, after  taking account of the differences between lattice and
continuum regularisation schemes. Moreover $f_K$ can be evaluated too
and one finds  that $f$ increases with increasing quark mass, just as
known from experiment. This increase is evidence for a contribution from
beyond the lowest order in chiral perturbation theory.

For the $\eta$ and $\eta'$, however, the flavour singlet combination 
will have gluonic contributions from the anomaly since
 \be
  \partial^{\mu} A^{\mu} = 2 m_q P + 2 N_f Q 
 \ee
 for $N_f$ quarks of mass $m_q$, where $P$ is the local bilinear
representing the  pseudoscalar current and $Q$ is known as the
topological charge density  and is given in terms of the gluonic field
by $Q=g^2 \epsilon_{\mu \nu \rho \sigma} F^{\mu \nu} F^{\rho \sigma}/(32
\pi^2)$. This identity underlines the inevitability of gluonic
contributions  to the flavour singlet meson interactions. 
 Thus for the flavour singlet (here labelled $\eta'$) defining a 
decay constant by
 \be
 \langle 0 | A^{\mu}| \eta' \rangle = f_{\eta'} q^{\mu}
 \label{pcaceta}
 \ee
 is less satisfactory since $f_{\eta'}$ will be scale dependent because
of  the gluonic contributions from the axial anomaly: namely there will 
be non-zero amplitudes 
 \be
 \langle 0 | Q | \eta' \rangle
 \label{etaglue}
 \ee

 I now address the issue of determining these decay constants directly
from QCD using lattice methods.  The study uses 2 flavours of degenerate
quark  and one defines the decay constants by eq.~\ref{pcacpi},
\ref{pcaceta}.
 For the isospin 1 state ($\pi$ - like), this is on a firm  footing
because of the axial ward identity hence $f_{\pi}$  will be scale
invariant. For the flavour singlet pseudoscalar meson, here called 
$\eta'$, the decay constant defined as in eq.~\ref{pcaceta} will not  be
scale invariant because of gluonic contributions arising from  the
anomaly~\cite{chpt} as discussed above. In an exploratory lattice 
study~\cite{cmcmeta} the decay constants are obtained with lattice
regularisation and one can   compare  the singlet and non-singlet
values.
 
 These decay constants can be thought of as giving the quark wave
function at the origin of the pseudoscalar meson. In principle it would
be  possible to explore also the local gluonic contributions to the 
flavour-singlet mesons (eq.~\ref{etaglue}) but it will be  difficult to
relate lattice regulated results to the continuum  and this has
not been attempted. 

 Since the mass splitting  between singlet and non-singlet is not
reproduced directly in quenched QCD, it  is  essential to use lattice
studies that  do include sea  quark effects in this study of decay
matrix elements.
 Results have been  obtained from lattices with   $N_f=2$  flavours of
degenerate sea quark. In this case there is no need for mixing angles 
to describe the spectrum (which will have one isoscalar and one
isovector  neutral particle). As a first crosscheck, one finds that
lattice studies  correctly give a flavour non-singlet decay constant
that increases with  quark mass in reasonable agreement  with
experiment~\cite{pdg} assuming a steady increase from $f_{nn}=131$ MeV
and  $f_{ns}=160$ MeV  to $f_{ss}$.

For the flavour singlet case,   the determinations of $f$  have
relatively  large statistical errors and the systematic error from
changing the type of fit is also comparable. For the case with $N_f=2$
degenerate quarks, the comparison of the flavour singlet  and
non-singlet shows that the singlet decay constants appear to be somewhat
larger, though the  errors are too big to substantiate this.

 Combining the mass dependence one finds in the flavour non-singlet sector
with  the near equality of singlet and non-singlet decay constants, we 
can deduce properties of the physical case with three light quarks. 
Thus,  in terms of the traditional treatment~\cite{pdg}, we would   expect
$f_{\eta}/f_{\pi} > 1$ and $f_{\eta'}/f_{\pi} > 1$. One way to minimise
the effects of mixing is to consider $X =(a_{\eta}^2 +
a_{\eta'}^2)/a_{\pi}^2$ where $a$ refers to the reduced decay amplitude.
Using the conventional formulae for the decay  amplitudes would then
give a value of $X = 3 r^2$ (where $r$ is a suitably weighted
average of  $f_{\eta}/f_{\pi} $ and $f_{\eta'}/f_{\pi}$ which are both
greater than 1). Thus the traditional treatment gives $X >3$ which is 
significantly larger than  the experimental value~\cite{pdg} of
2.64(24).  Thus it appears unlikely that the traditional treatment
(with the decay to $\gamma \gamma$  being given by the analogue of the
formula for pions) is correct for any mixing angle. 
 
 I conclude that there is no support for the traditional assumption
that the singlet decays are given by  a similar expression to the
non-singlet. As has been pointed out by many authors~\cite{chpt}, this 
is  plausible for at least two reasons: (i) the $\eta$ and $\eta'$
mesons are heavier and therefore  less likely to dominate the axial
current or, equivalently, higher order corrections to chiral
perturbation theory will be more important (ii) the flavour-singlet
axial anomaly has a gluonic component which will give additional
contributions to any  hadronic process.

\subsection{Where does the $\eta'$ mass come from?}

 The lattice studies show that the gluonic contributions that  build up
the disconnected correlators are such as to reproduce the  experimental
$\eta$ and $\eta'$ masses. This is of course as it should be -  QCD is
expected to be accurate for these phenomena - and it is the lattice
methodology that is being  tested. 

 One can now ask why these OZI-rule violating gluonic interactions are
substantial for  pseudoscalar mesons but small for most other cases
(scalar mesons excluded). 

 The culprit must be gluonic contributions with the required quantum
numbers. One possibility is a glueball, but the pseudoscalar glueball is
known~\cite{ukqcdgb}  to be heavier than 2 GeV and  would contribute
weakly and with the wrong sign (making the flavour singlet $q \bar{q}$
state  lighter than the non-singlet).  This glueball option is indeed 
appropriate for a discussion of the OZI violating effects for scalar
mesons. 

  For the pseudoscalar case, the presence of vacuum fluctuations with
those  quantum numbers is the candidate. These are commonly called
topological charge density fluctuations and they are necessary since 
the anomaly implies that $\epsilon_{\mu \nu \rho \sigma} F^{\mu \nu}
F^{\rho \sigma}$ is coupled to currents with pseudoscalar quantum
numbers. It must be emphasised that  these contributions  do not need to
have anything to do with isolated instantons.

 This relationship between fluctuations in topological charge density and 
the singlet mass generation can be made semi-quantitative via the 
Witten-Veneziano formula:
 \be
  m^2_{\eta'} = { 2 N_f \over f_{\pi}^2} \chi
 \ee
 where $\chi=\int d^4x <Q(x)Q(0)>$ is the topological susceptibility in
the pure gauge (Yang-Mills or quenched)  theory (here $\eta'$ is to be
interpreted as  the flavour singlet  component in the chiral limit). The
derivation of this  relationship (see ref~\cite{u1a} for a recent
discussion) involves studying correlations of topological charge density
assuming that $u=N_f/N_c$ can be varied. Then the  $u=0$ limit (where a
massless  contribution is present in the flavour-singlet channel in the
quenched chiral limit) can be  related to full QCD at $u \ne 0$ where 
there is no such massless flavour-singlet particle, hence  the
topological susceptibility is zero in the chiral limit. The relationship
follows from assuming that the lightest  flavour singlet pseudoscalar
meson dominates the behaviour of topological charge correlations in this 
region of small $u$ and near the chiral limit.

 Lattice techniques can be used to evaluate the quenched topological
susceptibility  and the value obtained is consistent~\cite{gusken,MQA} 
with this relationship. Evidence is also
mounting~\cite{durr,ukqcdtop,kovacs} that  $\chi$ does vanish in full
QCD as the sea quark mass goes to zero, as implied in the above
discussion.

 On a lattice one can investigate whether these topological charge
density fluctuations  arise from instantons or not, the latter being the
expectation  at large $N_c$. The Witten-Veneziano formula is a large
$N_c$  result and does seem to be qualitatively satisfied, which is some
evidence to suggest that the physical case with $N_c=3$  may be
qualitatively similar to the large $N_c$ world. The nature of the
topological charge density  on the lattice can be revealed by low lying
eigenmodes of the fermion operator, see ref.~\cite{lat01}, and this
suggests that  some elements of an instanton description are indeed
present. Varying $N_c$ in this study does suggest that at larger $N_c$
(e.g. 4)  the instanton features are less evident, as expected.  This
characterisation of the relationship between the topological charge density 
observed and classical tunnelling (instantons) is still not finally 
resolved.

\section{Conclusion}

 Flavour singlet mesons can, in principle, be easily studied on a
lattice.  In practice, lattice studies have been hampered by two
constraints. One is that the  disconnected quark diagrams needed for a
study of singlet mesons are intrinsically noisy.  Much larger data sets
(tens of thousands of gauge configurations) will be needed  to increase
precision. Another constraint is that one is unable to work with sea
quarks  substantially lighter than strange. Also it has not yet been
possible to  attempt a continuum limit extrapolation of the lattice
results,  although  a lattice formalism that should improve this
extrapolation was used. 

 A lattice study of the disconnected diagrams that cause flavour-singlet mesons
to differ from non-singlet mesons shows that these OZI-rule violating effects 
are only large for pseudoscalar and scalar mesons.  
Here we discuss the  structure of the mixing in the singlet pseudoscalar
mesons. 

 From the careful non-perturbative study of mass formulae for flavour
non-singlet pseudoscalar mesons  made of different quarks, one  deduces
that the  $ss$ state lies at 695 MeV. One then determines the pattern of
mixing for the  flavour singlet sector, obtaining $x_{ss} \approx 0.12$
GeV$^2$, $x_{nn}/x_{ss} \approx 2$ and  $x_{ns}^2 \approx x_{nn}
x_{ss}$. These conditions are indeed consistent and point to  a  mixing
close to maximal ($\phi=45 \pm 2^0$) in the $nn$, $ss$ basis (this
corresponds to a conventional ($\eta_8$, $\eta_1$) mixing $\theta$  of
$-10 \pm 2^0$). 

It is now  possible to explore the decay constants which give the
coupling to  a local quark antiquark pair for singlet pseudoscalar
mesons from the lattice. The results show similar decay constants for
singlet and non-singlet states of the same mass but with quite large
errors. This, combined with experimental data,  suggests that the
traditional description of $\gamma \gamma$ decays  is inadequate. There
will indeed be gluonic contributions  to singlet meson decays and these have
not yet been explored in detail from the lattice.

 Lattice studies  confirm the qualitative relationship between the
pseudoscalar singlet mass and the topological charge density
fluctuations.

Since  lattice techniques give a reasonable description of the 
flavour singlet  pseudoscalar mesons, it will be feasible to use such
techniques to study hadronic decays involving $\eta$ and $\eta'$.

\section{Acknowledgements}

 I thank Craig McNeile for many useful discussions and contributions.

\end{document}